\documentclass[american,12pt]{article}
\usepackage{pdfpages}
\usepackage{setspace}
\usepackage{tikz}
\usepackage{epsfig}
\usepackage{amsmath}
\usepackage{amsfonts}
\usepackage{pst-plot}
\usepackage{hyperref}
\usepackage{amssymb}
\usepackage{pgffor}
\usepackage[left=1.24in, right=1.24in, top=1.15in, bottom=1.15in, includefoot, headheight=1.2in]{geometry}
\usepackage{babel}
\usepackage{subfig}
\usepackage{tikz}
\usepackage{booktabs,siunitx}
\usepackage{accents}
\usepackage{mathrsfs}
\usepackage{amsfonts}
\usepackage{amssymb}
\usepackage{pgffor}
\usepackage{eufrak}
\usepackage[utf8]{inputenc}
\usepackage{enumitem}
\usepackage{booktabs}
\usepackage{diagbox}
\usepackage[authoryear]{natbib}
\usetikzlibrary{3d}
\usepackage{pgfplots}
\usepackage{enumitem}
\usetikzlibrary{positioning}
\usetikzlibrary{decorations.pathmorphing}
\usetikzlibrary{intersections, pgfplots.fillbetween}
\usetikzlibrary{matrix}
\usepackage{tikz,tikz-3dplot}
\tdplotsetmaincoords{60}{120}
\usepackage{qtree}
\usepackage{tikz-qtree}
\usetikzlibrary{trees}
\usepackage{scalerel}
\usepackage[12pt]{moresize}
\usepackage{amsmath,fouriernc}

\definecolor{mondrianRed}{rgb}{0.8666,0.1334,0}
\definecolor{mondrianBlue}{rgb}{0.133333, 0.313725, 0.584314}
\definecolor{mondrianYellow}{rgb}{0.980392, 0.788235, 0.00392157}
\definecolor{mondrianGrey}{rgb}{0.756863, 0.784314, 0.788235}
\definecolor{mondrianOrange}{rgb}{0.923529, 0.396078, 0.00196078}
\definecolor{mondrianGreen}{rgb}{0.556863, 0.54902, 0.294118}
\definecolor{mondrianCyan}{rgb}{0.345098, 0.431373, 0.439216}
\definecolor{mondrianWhite}{rgb}{0.976471, 0.976471, 0.976471}
\definecolor{mondrianPurple}{rgb}{0.107843, 0.158824, 0.292157}

\newcommand*{\field}[1]{\mathbb{#1}}

\usetikzlibrary {positioning}

\definecolor {processblue}{cmyk}{0.96,0,0,0}
\usetikzlibrary{arrows}
\usepackage[authoryear]{natbib}
\input{glyphtounicode}
\DeclareMathAlphabet{\mathpzc}{OT1}{pzc}{m}{it}

\makeatletter
\renewcommand{\section}{\@startsection{section}{1}{0mm}{-1.5\baselineskip}{0.8\baselineskip}{\normalfont\large\centering}}
\renewcommand{\subsection}{\@startsection{subsection}{2}{0mm}{-0.1\baselineskip}{0.5\baselineskip}{\normalfont\bf\flushleft}}
\renewcommand{\@seccntformat}[1]{\csname the#1\endcsname \hspace{+0mm}\large{.}\hspace{+1.9mm}}
\renewcommand{\@seccntformat}[2]{\csname the#1\endcsname \hspace{+0mm}\large{.}\hspace{+1.9mm}}
\makeatother

\newtheorem{theorem}{Theorem}

\newtheorem{corollary}{Corollary}

\newtheorem{example}{Example}

\newtheorem{lemma}{Lemma}

\renewcommand{\theequation}{\arabic{equation}}

\newlength{\extraspace}
\newlength{\extraspaces}
\newcounter{dummy}

\definecolor{mondrianRed}{rgb}{0.8666,0.1334,0}
\definecolor{mondrianBlue}{rgb}{0.133333, 0.313725, 0.584314}
\definecolor{mondrianYellow}{rgb}{0.980392, 0.788235, 0.00392157}
\definecolor{mondrianGrey}{rgb}{0.756863, 0.784314, 0.788235}
\definecolor{mondrianOrange}{rgb}{0.923529, 0.396078, 0.00196078}
\definecolor{mondrianGreen}{rgb}{0.556863, 0.54902, 0.294118}
\definecolor{mondrianCyan}{rgb}{0.345098, 0.431373, 0.439216}
\definecolor{mondrianWhite}{rgb}{0.976471, 0.976471, 0.976471}
\definecolor{mondrianPurple}{rgb}{0.107843, 0.158824, 0.292157}

\newcommand{\baa}{
\addtocounter{equation}{1} \setcounter{dummy}{\value{equation}}
\setcounter{equation}{0}
\renewcommand{\theequation}{\arabic{dummy}\alph{equation}}
\begin{eqnarray}
\addtolength{\abovedisplayskip}{\extraspaces}
\addtolength{\belowdisplayskip}{\extraspaces}
\addtolength{\abovedisplayshortskip}{\extraspace}
\addtolength{\belowdisplayshortskip}{\extraspace}}

\newcommand{\eaa}{
\end{eqnarray}
\setcounter{equation}{\value{dummy}}
\renewcommand{\theequation}{\arabic{equation}}}

\newcommand{\be}{\begin{equation}
\addtolength{\abovedisplayskip}{\extraspaces}
\addtolength{\belowdisplayskip}{\extraspaces}
\addtolength{\abovedisplayshortskip}{\extraspace}
\addtolength{\belowdisplayshortskip}{\extraspace}}
\newcommand{\ee}{\end{equation}}

\newcommand{\ba}{\begin{eqnarray}
\addtolength{\abovedisplayskip}{\extraspaces}
\addtolength{\belowdisplayskip}{\extraspaces}
\addtolength{\abovedisplayshortskip}{\extraspace}
\addtolength{\belowdisplayshortskip}{\extraspace}}
\newcommand{\ea}{\end{eqnarray}}

\newcommand{\bd}{\begin{displaymath}
\addtolength{\abovedisplayskip}{\extraspaces}
\addtolength{\belowdisplayskip}{\extraspaces}
\addtolength{\abovedisplayshortskip}{\extraspace}
\addtolength{\belowdisplayshortskip}{\extraspace}}
\newcommand{\ed}{\end{displaymath}}

\newcommand{\deel}[2]{{\textstyle{#1 \over #2}}}
\newcommand{\hf}{{\textstyle{1\over 2}}}

\def\inbar{\,\vrule height1.5ex width.4pt depth0pt}
\font\rms=cmr12 at 12pt
\def\ce{\relax\ifmmode\mathchoice
{\hbox{$\inbar\kern-.3em{\rm C}$}} {\hbox{$\inbar\kern-.3em{\rm
C}$}} {\lower.9pt\hbox{\rms $\inbar\kern-.3em{\rm C}$}}
{\lower1.2pt\hbox{\rms $\inbar\kern-.3em{\rm C}$}}
\else{$\inbar\kern-.3em{\rm C}$}\fi}
\font\cmss=cmss12 \font\cmsss=cmss12 at 12pt
\def\ze{\relax\ifmmode\mathchoice
{\hbox{\cmss Z\kern-.4em Z}}{\hbox{\cmss Z\kern-.4em Z}}
{\lower.9pt\hbox{\cmsss Z\kern-.4em Z}} {\lower1.2pt\hbox{\cmsss
Z\kern-.4em Z}}\else{\cmss Z\kern-.4em Z}\fi}

\newcommand{\refsection}[1]{
\vspace{1mm} \pagebreak[3] \addtocounter{section}{1}
\begin{center}
{\large #1}
\end{center}
\nopagebreak
\medskip
\nopagebreak}

\def\thebibliography#1{\refsection{\bf References}
\vspace*{-8mm}\list
 {\relax}{\itemsep=1pt \parsep=0pt
 \usecounter{enumiv}\leftmargin=3em\itemindent=-\leftmargin}%
 \def\newblock{\hskip .11em plus .33em minus .07em}
 \sloppy\clubpenalty4000\widowpenalty4000
 \sfcode`\.=1000\relax}

\newcommand{\startappendix}{
\renewcommand{\thesection}{\Alph{section}}
\setcounter{section}{0}
\renewcommand{\theequation}{\thesection.\arabic{equation}}}

\begin{document}

\setcounter{page}{0}
\thispagestyle{empty}

\begin{center}
{\Huge\sc The Economy's Potential}\\[3mm]{\LARGE\sc Duality and Equilibrium}\\[10mm]
{\large Jacob K. Goeree}\footnote{AGORA Center for Market Design, UNSW. I gratefully acknowledge funding from the Australian Research Council (DP190103888 and DP220102893). I thank Brett Williams for useful comments.}\\[5mm]
\today\\[16mm]
{\bf Abstract}
\end{center}
\addtolength{\baselineskip}{-1.2mm}
\vspace*{-3mm}

\noindent I introduce a concave function of allocations and prices -- the economy's potential -- which measures the difference between utilitarian social welfare and its dual. I show that Walrasian equilibria correspond to roots of the potential: allocations maximize weighted utility and prices minimize weighted indirect utility. Walrasian prices are ``utility clearing'' in the sense that the utilities consumers expect at Walrasian prices are just feasible. I discuss the implications of this simple duality for equilibrium existence, the welfare theorems, and the interpretation of Walrasian prices.

\vfill

\addtocounter{footnote}{-1}
\newpage

\addtolength{\baselineskip}{1.4mm}

\section{Introduction}

A landmark of economic theory concerns the determination of prices for all goods in the economy. General equilibrium theory originated with \mbox{Walras} ``\mbox{$\ldots$ whose} \mbox{system} of equations, defining equilibria in a system of interdependent quantities, is the Magna Carta of economic theory'' \citep{Schumpeter1954}. The modern approach to general equilibrium theory is due to \cite{ArrowDebrue1954} and \cite{McKenzie1954} who established existence of Walrasian equilibria. \cite{DuffieSonnenschein1989} describe the history and \mbox{importance} of this existence proof, which allowed general equilibrium theory to gain the central role it now occupies in economics and finance.

Arrow and Debreu's proof pertains to ``abstract economies'' or ``generalized games'' and builds on \citeauthor{nash1950}'s (\citeyear{nash1950}) existence proof for normal-form games. Their abstract approach establishes existence of a solution to a system of equations but does not reveal if, or why, it has desirable properties. This is the content of the welfare theorems, the modern versions of which are also due to \cite{Arrow1951} and \cite{Debreu1951}. The first welfare theorem -- that the price system results in a Pareto optimal allocation of resources -- is perhaps the central result in price \mbox{theory}.
The second welfare theorem states the converse, i.e. that every Pareto optimal allocation is Walrasian.

Despite their dual formulations, the proofs of the two welfare theorems are very different in nature. The first welfare theorem requires only positive marginal utility of income while the second welfare theorem hinges on the assumption of convexity. \cite{DuffieSonnenschein1989} note that as a result of Arrow and Debreu's work \mbox{``$\ldots$ the separateness of the two welfare theorems was brought into sharp focus.''}

I demonstrate that the welfare theorems encapsulate a simple \mbox{duality} property of Walrasian equilibria. To this end, I \mbox{introduce} the \textit{economy's} \textit{\mbox{potential}}, which is a weighted sum of consumers' utilities of their allocations minus their indirect utilities at given prices. I show that, for any welfare weights, the potential is a non-positive and strictly concave function with a unique root corresponding to the Walrasian equilibrium. The intuition is that allocations solve the primal problem of maximizing weighted utilities and prices solve the dual problem of minimizing weighted indirect utilities. Walrasian prices are ``utility clearing'' in that the utilities consumers expect are just feasible, i.e. the potential vanishes at the Walrasian equilibrium.

Usually, the economy is parameterized by initial endowments rather than welfare weights. While the set of endowments is of higher dimension than the set of welfare weights,\footnote{With $N>1$ consumers and $K>1$ goods the set of endowments is $NK$ dimensional while the set of welfare weights is $N$ dimensional.} no additional equilibria are introduced. The reason is that any endowments and resulting Walrasian price imply some set of welfare weights. Moreover, existence of Walrasian equilibrium for arbitrary endowments follows if there is at least one vector of ``equilibrium weights'' that produce the correct incomes. I show that existence of such equilibrium weights readily follows from the Poincare--Hopf theorem.

The next section presents a graphical illustration of duality for a simple exchange economy. Section 3 introduces the economy's potential, generalizes the duality result to any exchange economy, and shows that existence and the welfare theorems are direct corollaries of duality. Furthermore, duality provides a novel interpretation of the Walrasian price vector as the gradient of utilitarian social welfare.  Section 4 discusses possible applications of the potential to non-convex economies. Proofs can be found in the Appendix.

\section{An Example}
\label{sec:example}

\begin{figure}
\begin{center}
\begin{tikzpicture}[scale=0.65]
\draw[->] (-4,0) -- (6,0) node[right] {$u_1$};
\draw[->] (0,-4) -- (0,6) node[above] {$u_2$};
\draw [name path=A,red,very thick,  domain=0.1:0.9, samples=100] plot ({2*ln(6*\x/(1+\x))+ln(3*\x/(2-\x))}, {ln(3*(1-\x)/(1+\x))+2*ln(6*(1-\x)/(2-\x))});
\draw [name path=B] (-3.0581,-3.0581);
\tikzfillbetween[of=A and B]{gray, opacity=.2};
\draw [name path=C,blue,very thick, domain=0.1:0.9, samples=100] plot ({2*ln(2/(5*\x))+ln(1/(5*(1-\x)))}, {ln(4/(5*\x))+2*ln(8/(5*(1-\x)))});
\draw [name path=C,blue,very thick, domain=0.1:0.9, samples=100] plot ({2*ln(1/(\x))+ln(1/(2*(1-\x)))}, {ln(1/(2*\x))+2*ln(1/((1-\x)))});
\draw [name path=C,blue,very thick, domain=0.1:0.9, samples=100] plot ({2*ln(8/(5*\x))+ln(4/(5*(1-\x)))}, {ln(1/(5*\x))+2*ln(2/(5*(1-\x)))});
\node[scale=0.9,blue] at (6,-1.4) {$\alpha=\deel{4}{5}$};
\node[scale=0.9,blue] at (5,2) {$\alpha=\deel{1}{2}$};
\node[scale=0.9,blue] at (-2.2,5) {$\alpha=\deel{1}{5}$};
\draw [thin,dash dot,black] (1.38629-1,1.38629+1) -- (1.38629+1,1.38629-1);
\draw [thin,dotted,black] (1.38629-1+0.3,1.38629+1+0.3) -- (1.38629+1+0.3,1.38629-1+0.3);
\draw [thin,dashed,black] (1.38629-1-0.3,1.38629+1-0.3) -- (1.38629+1-0.3,1.38629-1-0.3);
\draw [thin,dotted,black] (1.38629-1+0.6,1.38629+1+0.6) -- (1.38629+1+0.6,1.38629-1+0.6);
\draw [thin,dashed,black] (1.38629-1-0.6,1.38629+1-0.6) -- (1.38629+1-0.6,1.38629-1-0.6);
\draw[->] [thin,solid,black] (1.38629-0.6,1.38629-0.6) -- (1.38629-0.1,1.38629-0.1);
\draw[->] [thin,solid,black] (1.38629+0.6,1.38629+0.6) -- (1.38629+0.1,1.38629+0.1);
\end{tikzpicture}
\vspace*{-3mm}
\end{center}
\caption{Illustration of duality. The shaded area is the utility possibility set and the red curve its frontier. The blue curves depict indirect utilities for given welfare weights as functions of prices. For $\alpha=\hf$, weighted utility is constant on the dashed lines and increases to the North-East while weighted indirect utility is constant on the dotted lines and decreases to the South-West. The unique point where weighted utility and weighted indirect utility match corresponds to the Walrasian equilibrium.}\label{fig:ex1}
\vspace*{-1mm}
\end{figure}
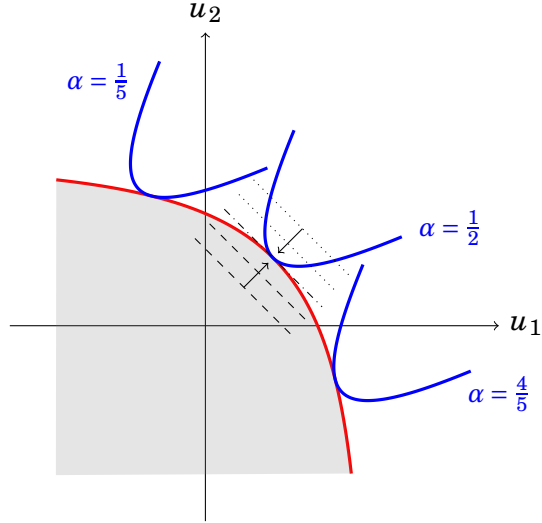

Consider an exchange economy with two consumers with Cobb-Douglas preferences $u_1(x,y)=2\log(x)+\log(y)$ and $u_2(x,y)=\log(x)+2\log(y)$. Suppose there are three units of each good in the economy. Welfare maximization
\begin{equation}\label{Wmax}
  \max_{{0\,\leq\,x_1+x_2\,\leq\,3}\atop{0\,\leq\,y_1+y_2\,\leq\,3}}\,\alpha u_1(x_1,y_1)+(1-\alpha)u_2(x_2,y_2)
\end{equation}
yields the Pareto-optimal allocations $(x_1(\alpha),y_1(\alpha))=(\deel{6\alpha}{1+\alpha},\deel{3\alpha}{2-\alpha})$ and $(x_2(\alpha),y_2(\alpha))=(\deel{3-3\alpha}{1+\alpha},\deel{6-6\alpha}{2-\alpha})$. The red curve in \mbox{Figure} \ref{fig:ex1} shows the resulting utility pairs $(u_1(\alpha),u_2(\alpha))$ for $\alpha\in(0,1)$. The shaded area corresponds to the utility possibility set.

The dual of \eqref{Wmax} entails minimizing a weighted sum of indirect utilities with respect to prices. If prices are normalized to sum to one, i.e. the price vector is $(p,1-p)$, then the indirect utility functions can be written as $v_1(p,m_1)=2\log(\deel{2m_1}{3p})+\log(\deel{m_1}{3(1-p)})$ and $v_2(p,m_2)=\log(\deel{m_2}{3p})+2\log(\deel{2m_2}{3(1-p)})$. The economy's total income is $3p+3(1-p)=3$ and a consumer's welfare weight determines her share:\footnote{More generally, a consumer's welfare weight is equal to the inverse of her marginal utility of income, see Theorem \ref{FDT} below. For the economy studied here this yields $\alpha=m_1/3$ and $1-\alpha=m_2/3$.} $m_1=3\alpha$ and $m_2=3(1-\alpha)$. The dual of \eqref{Wmax} is thus
\begin{equation}\label{Vmin}
  \min_{{0\,\leq\,p\,\leq\,1}}\,\alpha v_1(p,3\alpha)+(1-\alpha)v_2(p,3(1-\alpha))
\end{equation}
The blue curves show the indirect utility pairs $(v_1(p,3\alpha),v_2(p,3(1-\alpha))$ for different values of $\alpha$ and $p\in(0,1)$. For $\alpha=\hf$, the weighted sum of utilities is constant on the dashed lines and increasing in the North-East direction. The weighted sum of indirect utilities is constant on the dotted lines and decreasing in the South-West direction. There is a unique point where weighted utility is maximized and weighted indirect utility minimized and their values are equal. This point corresponds to the Walrasian equilibrium.

To see this, note that the solution to \eqref{Vmin} is $p(\alpha)=\deel{1}{3}(1+\alpha)$ and it is readily verified that the price ratio
\begin{displaymath}
\frac{p(\alpha)}{1-p(\alpha)}\,=\,\frac{1+\alpha}{2-\alpha}
\end{displaymath}
is equal to consumers' marginal rates of substitution at $(x_1(\alpha),y_1(\alpha))$ and $(x_2(\alpha),y_2(\alpha))$. Hence, these allocations maximize consumers' utilities given prices $(p(\alpha),1-p(\alpha))$. In other words, the Pareto-optimal allocations that follow from \eqref{Wmax} form a Walrasian equilibrium together with the price that follows from the dual program in \eqref{Vmin}.

\section{The Economy's Potential}
\label{sec:P_thm}

\noindent Consider an exchange economy with $\mathcal{N}=\{1,\ldots,N\}$ consumers and $\mathcal{K}=\{1,\ldots,K\}$ goods. For $i\in\mathcal{N}$, let $u_i:\field{R}^K_{\geq 0}\rightarrow\field{R}$ denote consumer $i$'s utility function and \mbox{$\omega_i\in\field{R}^K_{>0}$} consumer $i$'s endowment. I assume the utility functions are strictly increasing, strictly concave, and differentiable.\footnote{The usual assumption is that the utility functions are quasi-concave, which can be made concave by a monotone transformation. Intuitively, this construction implies that among utility functions with the same indifference curves there is one with non-increasing marginal utility. Concave functions are differentiable almost everywhere. For ease of presentation, I assume differentiability everywhere.}
For $k\in\mathcal{K}$, let $w_k=\sum_{i\in\mathcal{N}}\omega_{ik}$ denote the total amount of good $k$ in the economy.
The set of feasible allocations is
\begin{displaymath}
  F(w)\,=\,\{x\in\field{R}_{\geq0}^{NK}\,|\,\sum_{i\,\in\,\mathcal{N}}x_{ik}\,\leq\,w_k\,\,\,\forall\,k\in\mathcal{K}\}
\end{displaymath}
 For vectors $v,v'\in\field{R}^K$ let $\langle v|v'\rangle=\sum_{k\in\mathcal{K}}v_kv'_k$ denote the usual inner product. The Fenchel dual of $u_i(x_i)$ is defined as
\begin{equation}\label{fenchel}
\overline{v}_i(p)\,=\,\max_{x_i\,\geq\,0}\,\,u_i(x_i)-\langle p|x_i\rangle
\end{equation}
which is a strictly convex function of prices. An envelope-theorem argument establishes that the solution to \eqref{fenchel} satisfies $\overline{x}_i(p)=-\nabla_p\overline{v}_i(p)$, which is a simplified version of Roy's identity. The solution $\overline{x}_i(p)$ is strictly decreasing in each price and further satisfies $\lim_{p_k\downarrow0}\overline{x}_{ik}(p)=\infty$ and $\lim_{p_k\rightarrow\infty}\overline{x}_{ik}(p)=0$.

The next lemma relates $\overline{v}_i(p)$ and $\overline{x}_i(p)$ to the traditional indirect utility $v_i(p,m_i)$ and the Marshallian demand $x_i(p,m_i)$ respectively.
\begin{lemma}\label{lemma:roy}
For $i\in\mathcal{N}$, let $\lambda_i(p,m_i)$ solve $\langle p|\overline{x}_i(\lambda_ip)\rangle=m_i$, then
\begin{eqnarray}\label{roy}
  v_i(p,m_i) &=& \lambda_i(p,m_i)m_i+\overline{v}_i(\lambda_i(p,m_i)p)\label{v}\\
  x_i(p,m_i) &=& \overline{x}_i(\lambda_i(p,m_i)p)\label{x}
\end{eqnarray}
Moreover, $\lambda_i(p,m_i)$ equals the marginal utility of income, $\lambda_i(p,m_i)=\partial v_i(p,m_i)/\partial m_i$, and for any price vector, $\lambda_i(p,m_i)$ is strictly positive and strictly decreasing in $m_i$ with $\lim_{m_i\downarrow 0}\lambda_i(p,m_i)=\infty$ and $\lim_{m_i\rightarrow\infty}\lambda_i(p,m_i)=0$.
\end{lemma}
\begin{example}[CES utility]\label{ex:CES}{\em
For $i\in\mathcal{N}$ and $\rho_i<1$, consider the CES utilities
\begin{displaymath}
  u_i(x_i)\,=\,\frac{1}{\rho_i}\,\log(\sum_{k\,\in\,\mathcal{K}}a_{ik}x_{ik}^{\rho_i})
\end{displaymath}
with Fenchel duals
\begin{displaymath}
  \overline{v}_i(p)\,=\,\frac{1-\rho_i}{\rho_i}\,\log\bigl(\,\sum_{k\,\in\,\mathcal{K}}a_{ik}\bigl(p_k/a_{ik})^{\frac{\rho_i}{\rho_i-1}}\bigr)-1
\end{displaymath}
Roy's identity yields
\begin{displaymath}
  \overline{x}_{ik}(p)\,=\,\frac{(p_k/a_{ik})^{\frac{1}{\rho_i-1}}}{\sum_{\ell\,\in\,\mathcal{K}}a_{i\ell}(p_\ell/a_{i\ell})^{\frac{\rho_i}{\rho_i-1}}}
\end{displaymath}
which satisfies $\overline{x}_{ik}(p/m_i)=m_i\overline{x}_{ik}(p)$ and $\langle p|\overline{x}_i(p)\rangle=1$ so $\lambda_i(p,m_i)=\deel{1}{m_i}$. \mbox{Marshallian} demand is $x_i(p,m_i)=m_i\overline{x}_i(p)$ and indirect utility is $v_i(p,m_i)=1+\overline{v}_i(p)+\log(m_i)$.
$\hfill\blacksquare$}
\end{example}
\noindent Let $U_\alpha(x)=\sum_{i\in\mathcal{N}}\alpha_iu_i(x_i)$ denote weighted aggregate utility for some positive weight vector $\alpha\in\field{R}^N_{>0}$.
The welfare-maximization program
\begin{equation}\label{weightedWelfare}
  W_\alpha(w)\,=\,\max_{x\,\in\,F(w)}\,U_\alpha(x)
\end{equation}
is equivalent to the saddle-point problem
\begin{displaymath}
  W_\alpha(w)\,=\,\min_{p\,\geq\,0}\,\,\max_{x\,\geq\,0}\,\sum_{i\,\in\,\mathcal{N}}\,\alpha_iu_i(x_i)-\langle p|x_i-\omega_i\rangle
\end{displaymath}
where the multiplier for the feasibility constraint $\sum_{i\in\mathcal{N}}(x_i-\omega_i)\leq0$ is denoted $p$ on purpose. The maximization over allocations is solved by
$\overline{x}_i(p/\alpha_i)$ and using the Fenchel duals $\overline{v}_i$ we can write the result as
\begin{equation}\label{Walpha}
  W_\alpha(w)\,=\,\min_{p\,\geq\,0}\,V_\alpha(p,w)
\end{equation}
where
\begin{equation}\label{Usol}
  V_\alpha(p,w)\,=\,\langle p|w\rangle+\sum_{i\,\in\,\mathcal{N}}\,\alpha_i\overline{v}_i(p/\alpha_i)
\end{equation}
is strictly convex in prices.

The \textit{economy's potential} is defined as
\begin{equation}\label{pot}
  Y_\alpha(x,p,w)\,=\,U_\alpha(x)-V_\alpha(p,w)
\end{equation}
From \eqref{weightedWelfare} and \eqref{Walpha} it follows that the potential is non-positive for all feasible allocations and non-negative prices. Moreover, it is a strictly concave function of allocations and prices since $U_\alpha(x)$ is strictly concave and $V_\alpha(p,w)$ is strictly convex.

Let $\lambda_i^{(-1)}(p,\cdot\,)$ denote the inverse of $\lambda_i(p,m_i)$ with respect to the income $m_i$.
\begin{theorem}\label{FDT}
If $(x,p)$ is a Walrasian equilibrium of the economy with incomes $m_i$ for $i\in\mathcal{N}$ then $\,Y_\alpha(x,p,w)=0$ for $\alpha_i=1/\lambda_i(p,m_i)$ and $i\in\mathcal{N}$.
\mbox{Conversely, for any $\alpha\in\field{R}^N_{>0}$,} $\,Y_\alpha(x,p,w)$ has a unique root $(x(\alpha),p(\alpha))$, which is the Walrasian equilibrium of the economy with incomes $m_i(\alpha)=\lambda_i^{(-1)}(p(\alpha),\frac{1}{\alpha_i})$ for $i\in\mathcal{N}$ and $\sum_{i\in\mathcal{N}}m_i(\alpha)=\langle p(\alpha)|w\rangle$.
\end{theorem}

\noindent Evaluated at the equilibrium price, \eqref{Usol} can also be expressed in terms of the standard indirect utilities
\begin{displaymath}
  V_\alpha(p(\alpha),w)\,=\,\sum_{i\,\in\,\mathcal{N}}\alpha_iv_i(p(\alpha),m_i(\alpha))
\end{displaymath}
which follows from \eqref{v} together with $\lambda_i(p(\alpha),m_i(\alpha))=\deel{1}{\alpha_i}$ and $\sum_{i\in\mathcal{N}}m_i(\alpha)=\langle p(\alpha)|w\rangle$.

\subsection{Welfare Theorems}

A necessary condition for the pair $(x,p)$ to be a root of the potential is that $x$ maximizes welfare, i.e. it is Pareto optimal. The first part of Theorem \ref{FDT} thus implies the first welfare theorem and the converse part implies the second welfare theorem.
\begin{corollary}\label{wt}
Theorem \ref{FDT} implies that any Walrasian allocation is Pareto-optimal and vice versa.
\end{corollary}

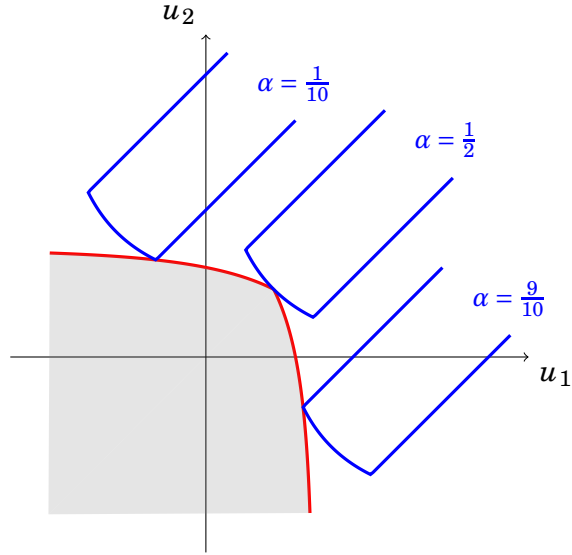
\begin{figure}
\begin{center}
\begin{tikzpicture}[scale=1.3]
\draw[->] (-2,0) -- (3.3,0) node[below right] {$u_1$};
\draw[->] (0,-2) -- (0,3.3) node[above left] {$u_2$};
\draw [name path=A,red,very thick,  domain=0:0.9, samples=100] plot ({ln(2+\x)}, {ln(2-2*\x)});
\draw [name path=B,red,very thick,  domain=0:0.9, samples=100] plot ({ln(2-2*\x)}, {ln(2+\x)});
\draw [name path=C] (-1.61,-1.61);
\tikzfillbetween[of=A and C]{gray, opacity=.2};
\tikzfillbetween[of=B and C]{gray, opacity=.2};
\draw [name path=D,blue,very thick, domain=0.08:0.92, samples=100] plot ({ln(1/10*max(2/\x,1/(1-\x))}, {ln(9/10*max(1/\x,2/(1-\x)))});
\draw [name path=D,blue,very thick, domain=0.08:0.92, samples=100] plot ({ln(1/2*max(2/\x,1/(1-\x))}, {ln(1/2*max(1/\x,2/(1-\x)))});
\draw [name path=D,blue,very thick, domain=0.08:0.92, samples=100] plot ({ln(9/10*max(2/\x,1/(1-\x))}, {ln(1/10*max(1/\x,2/(1-\x)))});

\node[scale=0.9,blue] at (3.1,0.6) {$\alpha=\deel{9}{10}$};
\node[scale=0.9,blue] at (2.45,2.22) {$\alpha=\deel{1}{2}$};
\node[scale=0.9,blue] at (0.9,2.8) {$\alpha=\deel{1}{10}$};
\end{tikzpicture}
\end{center}
\caption{Illustration of duality for the economy in Example \ref{ex2}.}
\label{fig:ex2}
\end{figure}

\noindent For the economy in Section \ref{sec:example}, the Walrasian equilibrium price and allocation can be obtained from either the primal welfare-maximization program or the dual program of minimizing weighted indirect utility. For instance, the primal program yields the allocations and the price follows from consumers' marginal rates of substitution. Alternatively, the dual program yields the price and the allocations follow from Roy's identity. This is a consequence of the strict concavity assumption.  If we relax this assumption to concavity, the tangency in Figure \ref{fig:ex1} does not necessarily occur. Nonetheless, Walrasian prices are ``utility clearing,'' i.e. they force a zero potential.
\begin{example}[Linear utility]\label{ex2}
{\em Consider an exchange economy with two consumers with linear preferences $u_1(x,y)=\log(2x+y)$ and $u_2(x,y)=\log(x+2y)$. Suppose there is one unit of each good in the economy. Welfare maximization yields the Pareto-optimal allocations
\begin{displaymath}
  (x_1(\alpha),y_1(\alpha))\,=\,\left\{
  \begin{array}{lll}(3\alpha,0) & \mbox{if} & \alpha\,\leq\,\frac{1}{3}\\ (1,0) & \mbox{if} & \frac{1}{3}\,\leq\,\alpha\,\leq\,\frac{2}{3} \\ (1,3\alpha-2) & \mbox{if} & \alpha\,\geq\,\frac{2}{3}\end{array}\right.
\end{displaymath}
and $(x_2(\alpha),y_2(\alpha))=(1-x_1(\alpha),1-y_1(\alpha))$. The frontier of the utility-possibility set, depicted by the red curve in Figure \ref{fig:ex2}, corresponds to the resulting utility pairs $(u_1,u_2)$.
The Pareto optimal allocations lie on the boundary of the Edgeworth box for any welfare weight and the Walrasian price does not follow from consumers' marginal rates of substitution. Instead it follows from mininizing $V_\alpha(p,w)$, which yields
\begin{displaymath}
  \frac{p(\alpha)}{1-p(\alpha)}\,=\,\left\{
  \begin{array}{lll}\frac{1}{2} & \mbox{if} & \alpha\,\leq\,\frac{1}{3}\\ \frac{\alpha}{1-\alpha} & \mbox{if} & \frac{1}{3}\,\leq\,\alpha\,\leq\,\frac{2}{3} \\ 2 & \mbox{if} & \alpha\,\geq\,\frac{2}{3}\end{array}\right.
\end{displaymath}
This outcome is illustrated by the blue curves in Figure \ref{fig:ex2}.
$\hfill\blacksquare$}
\end{example}

\subsection{The Greedy Invisible Hand}

An envelope-theorem argument applied to \eqref{Walpha} establishes the Walrasian price vector as the gradient of utilitarian social welfare.
\begin{corollary}\label{greedy}
If $(x,p)$ is a Walrasian equilibrium then
\begin{equation}\label{solution}
  p\,=\,\nabla_w W_{\alpha}(w)
\end{equation}
for $\alpha_i=1/\lambda_i(p,m_i)$ and $x_i=\overline{x}_i(\lambda_i(p,m_i)p)$ for $i\in\mathcal{N}$.
\end{corollary}
The characterization of the Walrasian equilibrium price as the gradient of utilitarian social welfare is surprising -- Adam Smith's ``invisible hand'' steers market participants to a state of greatest happiness in a simple greedy manner. Yet, the characterization is intuitive as it implies a balance of individual and social incentives:
\begin{equation}\label{MRS}
    \frac{\partial_kW_\alpha(w)}{\partial_\ell W_\alpha(w)}\,=\,\frac{\partial_ku_i(x_i)}{\partial_\ell u_i(x_i)}
\end{equation}
for $i\in\mathcal{N}$, $k,\ell\in\mathcal{K}$ and $\alpha_i=1/\lambda_i(p,m_i)$. In other words, individuals' marginal rates of substitution match the social marginal rate of substitution.
Corollary \ref{greedy} provides a simple-yet-powerful way to analytically characterize Walrasian equilibria.
\begin{example}[Homogeneous CES] \label{ex:CES}
{\em
Suppose there are $K$ goods and $N$ consumers with CES utilities
\begin{displaymath}
  u_i(x)\,=\,\frac{1}{\rho}\,\log\bigl(\,\sum_{k\,\in\,\mathcal{K}}\,a_kx_{k}^\rho\bigr)
\end{displaymath}
for $\rho<1$. Consumers' endowments are $\omega_i$ and the total endowments are $w=\sum_{i\in\mathcal{N}}\omega_i$. Aggregate utility $U_\alpha(x)=\sum_{i\in\mathcal{N}}\alpha_iu_i(x_i)$
is maximized at $x_i=(\alpha_i/\sum_{j\in\mathcal{N}}\alpha_j)w$. The social weights are $\alpha_i=1/(\partial v_i/\partial m_i)=m_i$, so the gradient of utilitarian social welfare is
\begin{displaymath}
  \nabla_w W_{\alpha}(w)\,=\,\frac{\sum_{i\,\in\,\mathcal{N}}m_i}{\sum_{k\,\in\,\mathcal{K}}\,w_k^{\rho}}\,\,w^{\rho-1}\,=\,p
\end{displaymath}
resulting in price ratios $p_k/p_\ell=(w_k/w_\ell)^\rho$.
Using $m_i=\langle p|\omega_i\rangle$ and $x_i=(\alpha_i/\sum_{j\in\mathcal{N}}\alpha_j)w$ yields the Walrasian equilibrium allocations
\begin{displaymath}
  x_i\,=\,\frac{\sum_{k\,\in\,\mathcal{K}}\,\omega_{ik}w_k^{\rho-1}}{\sum_{k\,\in\,\mathcal{K}}\,w_k^{\rho}}\,w
\end{displaymath}
for $i\in\mathcal{N}$. Note that I did not need to solve any individual consumer's maximization problem to derive the Walrasian equilibrium price and allocations.
$\hfill\blacksquare$}
\end{example}

\subsection{Existence of Walrasian Equilibria}

Theorem \ref{FDT} shows there is a unique Walrasian equilibrium for any welfare weights. No fixed-point arguments are needed. A simple duality result establishes the Walrasian equilibrium as the unique maximum, and root, of a strictly concave potential.

Usually, the economy is parameterized by endowments $\omega_i\in\field{R}_{>0}$ for $i\in\mathcal{N}$ rather than welfare weights. Theorem \ref{FDT} shows that if there is a Walrasian equilibrium $(x,p)$ then it is a root of the potential for weights that equal the inverse of the marginal utility of income: $\alpha_i=1/\lambda_i(p,\langle p|\omega_i\rangle)$. Hence, despite the set of endowments being of higher dimension $(NK)$ than the set of welfare weights $(N)$, no additional Walrasian equilibria are added when parameterizing the economy by endowments.

Do Walrasian equilibria exist for any endowments? \cite{ArrowDebrue1954} have answered this question affirmatively by extending \citeauthor{nash1950}'s (\citeyear{nash1950}) existence proof. Here I present a simpler argument by showing that for any initial endowments there are ``equilibrium weights'' that produces the correct incomes. Since $Y_{\kappa\alpha}(x,\kappa p,w)=\kappa Y_\alpha(x,p,w)$ for any $\kappa>0$ we can, without loss of generality, scale the weight vector $\alpha$ so its entries sum to one, i.e. $\alpha$ belongs to the simplex $\Sigma_N$. For $i\in\mathcal{N}$, consider the fixed-point equations
\begin{equation}\label{eqW}
 \langle p(\alpha)|\omega_i\rangle-m_i(\alpha)\,=\,0
\end{equation}
By Theorem 1, $\sum_{i\in\mathcal{N}}m_i(\alpha)=\langle p|w\rangle$, so the left side of \eqref{eqW} defines a vector field on $\Sigma_N$. Moreover, $\lim_{\alpha_i\downarrow 0}m_i(\alpha)=0$ by Lemma \ref{lemma:roy}, so the vector field points inward on the boundary of $\Sigma_N$. By the Poincare--Hopf theorem such a vector field has at least one zero in the interior of $\Sigma_N$.
\begin{corollary}\label{exist}
For any economy parametrized by endowments $\omega_i\in\field{R}_{>0}$, $i\in\mathcal{N}$ there exists a weight vector $\alpha\in\Sigma_N$ such that the root of $Y_\alpha(x,p,w)$ is a Walrasian equilibrium.
\end{corollary}

\noindent The solution to \eqref{eqW} is not necessarily unique as the next example shows.
\begin{example}\label{multi}{\em
Consider an exchange economy with two goods and two consumers with utilities $u_1(x,y)=\deel{3}{2}x^{\frac{2}{3}}-\hf y^{-2}$ and $u_2(x,y)=\deel{3}{2}y^{\frac{2}{3}}-\hf x^{-2}$ and endowments $\omega_1=(\deel{11}{6},\deel{1}{6})$ and $\omega_2=(\deel{1}{6},\deel{11}{6})$. The Fenchel duals are
\begin{displaymath}
  \overline{v}_1(p_1,p_2)\,=\,\frac{1}{2}\,\Bigl(\,\frac{1}{p_1^2}-3p_2^{\frac{2}{3}}\,\Bigr)
\end{displaymath}
and $\overline{v}_2(p_1,p_2)=\overline{v}_1(p_2,p_1)$.
The prices $(p_1(\alpha),p_2(\alpha))$ that minimize $V_\alpha(p,w)$ solve
\begin{displaymath}
  \Bigl(\frac{\alpha}{p_1(\alpha)}\Bigr)^{3}+\Bigl(\frac{1-\alpha}{p_1(\alpha)}\Bigr)^{\frac{1}{3}}\,=\,2
\end{displaymath}
and the equation for $p_2(\alpha)$ follows by interchanging $\alpha$ and $1-\alpha$, i.e. $p_2(\alpha)=p_1(1-\alpha)$. The incomes are $m_1(\alpha)=\alpha^3/p_1(\alpha)^2+\alpha^{\frac{1}{3}}p_2(\alpha)^{\frac{2}{3}}$ and $m_2(\alpha)=m_1(1-\alpha)$.  The fixed-point condition \eqref{eqW} for the equilibrium weight can be written as
\begin{displaymath}
  p_1(\alpha)\Bigl(\Bigl(\frac{1-\alpha}{p_1(\alpha)}\Bigr)^{\frac{1}{3}}-\frac{1}{6}\Bigr)\,=\,p_1(1-\alpha)\Bigl(\Bigl(\frac{\alpha}{p_1(1-\alpha)}\Bigr)^{\frac{1}{3}}-\frac{1}{6}\Bigr)
\end{displaymath}
An obvious solution is $\alpha=\hf$ and prices $(p_1,p_2)=(\hf,\hf)$. Two other solutions are, approximately, $\alpha\approx 0.09$, $(p_1,p_2)\approx(0.16,0.79)$ and $\alpha\approx 0.91$, $(p_1,p_2)\approx(0.79,0.16)$.
$\hfill\blacksquare$}
\end{example}

\noindent To summarize, parameterizing the economy using welfare weights is economical in two ways.  First, there is exactly one Walrasian equilibrium for every welfare weight. Second, this unique Walrasian equilibrium follows from duality rather than from a system of fixed-point conditions. In contrast, parameterizing the economy using endowments is \textit{un}economical for three reasons. First, if $(x,p)$ is a Walrasian equilibrium for the endowments $\omega_i$ then it is also a Walrasian equilibrium for any endowments $\omega'_i$ that satisfy $\langle p|\omega_i\rangle=\langle p|\omega'_i\rangle$ for $i\in\mathcal{N}$. Second, there may exist multiple Walrasian equilibria for some endowments as Example \ref{multi} shows. Third, computing Walrasian equilibria necessarily involves solving a system of fixed-point conditions, see \eqref{eqW}. It should be noted that, since the number of goods $(K)$ is typically assumed to be far larger than the number of consumers $(N)$, \eqref{eqW} provides a computationally efficient alternative to solving fixed-point conditions for the equilibrium prices.

Importantly, even if the economy is parameterized by endowments there are no additional Walrasian equilibria with possibly different features than the potential's roots. The Paretian--Walrasian duality derived above thus applies to \textit{all} Walrasian equilibria (even when computed using standard fixed-point conditions).

\section{Outlook}

The potential provides a litmus test for equilibrium existence. Given preferences and endowments it is a mechanical exercise to compute the potential's maximum value. A Walrasian equilibrium exists if and only if this exercise returns nil.

If not, general equilibrium theory is quiet about the allocations and prices that ensue even when there are obvious gains from trade. For instance, suppose two consumers have $\max(x,y)$ preferences and endowments $\omega_1=(2,2)$ and $\omega_2=(1,1)$. At any price vector $(p,1-p)$ consumer 1 demands at least four units of one of the goods, ruling out Walrasian equilibrium. Yet, consumers can exchange one unit of either good for one unit of the other and both be better off.

Non-existence of equilibrium usually stops any further analysis, but that does not mean that gains from trade will not be seized -- the economy continues to operate after all.
\cite{GoereeRoger2022} demonstrate that the outcome in which one unit is exchanged corresponds to a maximum of the potential, albeit not a root. They term these maxima ``$Y$ equilibria'' and show they exist in \textit{any} economy, including non-convex ones. As such, the potential complements general equilibrium's incomplete toolkit and provides a compass to navigate economics' terra incognita of non-convex economies.

\newpage
\startappendix
\setcounter{equation}{0}
\addtolength{\baselineskip}{-0.7mm}

\section{Proofs}
\label{AppA}

\noindent \textbf{Proof of Lemma 1}.
The indirect utility $v_i(p,m_i)$ follows from the saddle-point problem
\begin{equation}\label{indirect}
  v_i(p,m_i)\,=\,\min_{\lambda_i\,\geq\,0}\max_{x_i\,\geq\,0}\,u_i(x_i)-\lambda_i(\langle p|x_i\rangle-m_i)
\end{equation}
and is related to the Fenchel dual as follows
\begin{displaymath}
  v_i(p,m_i)\,=\,\min_{\lambda_i\,\geq\,0}\,\lambda_im_i+\overline{v}_i(\lambda_ip)
\end{displaymath}
Let $\lambda_i(p,m_i)$ be the solution to the minimization problem then
\begin{equation}\label{indirect2}
  v_i(p,m_i)\,=\,\lambda_i(p,m_i)m_i+\overline{v}_i(\lambda_i(p,m_i)p)
\end{equation}
where the $\lambda_i(p,m_i)$ for $i\in\mathcal{N}$ are such that budget constraints are binding:
\begin{equation}\label{FOClambdahat}
  \langle p|\overline{x}_i(\lambda_i(p,m_i)p)\rangle\,=\,m_i
\end{equation}
From \eqref{indirect}, $\nabla_p v_i(p,m_i)=\lambda_i(p,m_i)(\omega_i-x_i)$ since $m_i=\langle p|\omega_i\rangle$. In addition, from \eqref{indirect2}, $\nabla_p v_i(p,m_i)=\lambda_i(p,m_i)(\omega_i+\nabla_p\overline{v}_i(\lambda_i(p,m_i)p))$. Combining these results yields a simplified version of Roy's identity: $x_i(p,m_i)=\overline{x}_i(\lambda_i(p,m_i)p)=-\nabla_pv_i(\lambda_i(p,m_i)p)$.
From \eqref{indirect2} it follows that $\partial v_i/\partial m_i=\lambda_i(p,m_i)$. The budget constraints \eqref{FOClambdahat} imply that $\lambda_i(p,m_i)>0$ and that $\overline{x}_i(\lambda_i(p,m_i)p)$ is strictly decreasing in $p$ and strictly increasing in $m_i$. The latter implies that $\lambda_i(p,m_i)$ is strictly decreasing in $m_i$. From \eqref{FOClambdahat} it further follows that  $\overline{x}_i(\lambda_i(p,m_i)p)$ vanishes when $m_i=0$ and $\overline{x}_i(\lambda_i(p,m_i)p)$ diverges when $m_i=\infty$. Hence, $\lim_{m_i\downarrow0}\lambda_i(p,m_i)=\infty$ and $\lim_{m_i\rightarrow\infty}\lambda_i(p,m_i)=0$.
$\hfill\blacksquare$

\medskip\medskip

\noindent \textbf{Proof of Theorem 1}.
If $(x,p)$ is a Walrasian equilibrium the Marshallian demands satisfy $x_i=-\nabla\overline{v}_i(\lambda_i(p,m_i)p)$, see Lemma \ref{lemma:roy}. This can be inverted to $\lambda_i(p,m_i)p=\nabla u_i(x_i)$, see e.g. \citeauthor{Rockafellar1970} (\citeyear{Rockafellar1970}, Th. 26.5).\footnote{Theorem 26.5 in \cite{Rockafellar1970} relates the gradient of a strictly convex function to the gradient of its dual, which is also a strictly convex function. To apply the theorem use $-u$ and $\overline{v}$.} When $\alpha_i=1/\lambda_i(p,m_i)$ we thus have $p=\alpha_i\nabla u_i(x_i)$, so the $x_i$ satisfy the first-order conditions for maximizing $U_\alpha$. Hence, $x=\text{argmax}_{x'}U_\alpha(x')$ as $U_\alpha$ is strictly concave. The Walrasian price $p$ is market clearing so $0=\sum_{i\in\mathcal{N}}(\omega_i-x_i(p,m_i))=\nabla_p V_\alpha(p,w)$ when $\alpha_i=1/\lambda_i(p,m_i)$ for $i\in\mathcal{N}$. So $p$ satisfies the first-order condition for minimizing $V_\alpha$ and $p=\text{argmin}_{p'}V_\alpha(p',w)$ as $V_\alpha$ is strictly convex. Since $\max_{x}U_\alpha(x)=\min_{p}V_\alpha(p,w)$ the potential vanishes at $(x,p)$.

For the converse part, since $p(\alpha)$ minimizes $V_\alpha(p,w)$ it follows that $\nabla_p V_\alpha(p,w)=\sum_{i\in\mathcal{N}}(\omega_i-\overline{x}_i(p(\alpha)/\alpha_i))=0$, i.e. the $\overline{x}_i(p(\alpha)/\alpha_i)$ satisfy feasibility. The unique solution $x(\alpha)$ to $\max_{x\in F(w)}\,U_\alpha(x)$ satisfies $\alpha_i\nabla u_i(x_i(\alpha))=q(\alpha)$ for $i\in\mathcal{N}$ and some price vector $q(\alpha)$. This can be inverted as $x_i(\alpha)=-\nabla\overline{v}_i(q(\alpha)/\alpha_i)=\overline{x}_i(q(\alpha)/\alpha_i)$. Since $x(\alpha)$ is feasible by construction, $q(\alpha)$ also minimizes $V_\alpha(p,w)$. Strict convexity of $V_\alpha(p,w)$ implies $q(\alpha)=p(\alpha)$. By Lemma \ref{lemma:roy} the
$\overline{x}_i(p(\alpha)/\alpha_i)$ are optimal Marshallian \mbox{demands} at $p(\alpha)$ and incomes $m_i$ if $\alpha_i=1/\lambda_i(p(\alpha),m_i)$, which can be inverted as $m_i=\lambda_i^{(-1)}(p(\alpha),\deel{1}{\alpha_i})$. Taking the inner product of the feasibility condition with the price vector yields $\langle p(\alpha)|w\rangle=\sum_{i\in\mathcal{N}}\langle p(\alpha)|\overline{x}_i(p(\alpha)/\alpha_i)\rangle=\sum_{i\in\mathcal{N}}m_i(\alpha)$. To summarize, $(x(\alpha),p(\alpha))$ is a Walrasian equilibrium of the economy with incomes $m_i(\alpha)$.
 $\hfill\blacksquare$

\vspace*{9mm}
\addtolength{\baselineskip}{-0.3mm}
\bibliography{references}
\bibliographystyle{chicago}

\end{document}